\documentclass[11pt]{article}
\usepackage{axodraw}
\input pix.sty

\newcommand{\us}[1]{\frac{1}{d-x}}

\newcommand{\bmu}{\bar\mu}
\newcommand{\tinymsbar}{{\overline{\mbox{\tiny\rm{MS}}}}}
\def\Ada(#1,#2)(#3,#4,#5){\DashCArc(#1,#2)(#3,#4,#5){3}}
\def\Lda(#1,#2)(#3,#4){\DashLine(#1,#2)(#3,#4){3}}

\def\ToprVBblob(#1,#2,#3,#4){\picb{#1(30,15)(15,-120,120)%
 #2(30,15)(15,120,240) #3(15,15)(15,60,300) #4(15,15)(15,-60,60)%
 \GCirc(45,15){2}{0}}}

\newcommand{\Nf}{N_{\rm f}}
\newcommand{\Nc}{N_{\rm c}}
\newcommand{\rmO}{{\mathcal{O}}}

\renewcommand{\bG}{\beta_\rmi{G}}

\def\lsi{\raise0.3ex\hbox{$<$\kern-0.75em\raise-1.1ex\hbox{$\sim$}}}
\def\gsi{\raise0.3ex\hbox{$>$\kern-0.75em\raise-1.1ex\hbox{$\sim$}}}

\newcommand{\gsim}{\mathop{\gsi}}

\newcommand{\bfx}{{\bf x}}

\newcommand{\Tc}{T_{\rm c}}

\newcommand{\rmii}[1]{{\mbox{\tiny\rm{#1}}}}
\newcommand{\amE}{\hat m} 
\newcommand{\Lambdamsbar}{{\Lambda_\tinymsbar}}

\makeatletter \@addtoreset{equation}{section} \makeatother
\renewcommand{\theequation}{\arabic{section}.\arabic{equation}}
\makeatletter
\renewcommand\section{\@startsection {section}{1}{\z@}%
                                   {-5.5ex \@plus -1ex \@minus -.2ex}
                                   {2.3ex \@plus.2ex}%
                                   {\normalfont\large\bfseries}}
\renewcommand\subsection{\@startsection{subsection}{2}{\z@}%
                                     {-3.25ex\@plus -1ex \@minus -.2ex}%
                                     {1.5ex \@plus .2ex}%
                                     {\normalfont\normalsize\bfseries}}
\renewcommand\thesection {\@arabic\c@section}
\renewcommand\thesubsection   {\thesection.\@arabic\c@subsection}
\renewcommand{\@seccntformat}[1]{%
\csname the#1\endcsname.\hspace{1.0em}}
\makeatother

\begin{document}

\begin{titlepage}
\begin{flushright}
BI-TP 2008/26\\
\end{flushright}
\begin{centering}

\vfill
 
{\Large{\bf Three-dimensional physics and the pressure of hot QCD}}

\vspace{0.8cm}

A. Hietanen$^{\rm a}$, 
K. Kajantie$^{\rm b}$, 
M.~Laine$^{\rm c}$, 
K.~Rummukainen$^{\rm d}$, 
Y. Schr\"oder$^{\rm c}$ 

\vspace{0.8cm}

{\em $^{\rm a}$%
Department of Physics, 
Florida International University, Miami, FL 33199, USA\\}

\vspace{0.3cm}

{\em $^{\rm b}$%
Helsinki Institute of Physics,
P.O.Box 64, FIN-00014 University of Helsinki, Finland\\}

\vspace{0.3cm}

{\em $^{\rm c}$%
Faculty of Physics, University of Bielefeld, 
D-33501 Bielefeld, Germany\\}

\vspace{0.3cm}

{\em $^{\rm d}$%
Department of Physics, University of Oulu, 
P.O.Box 3000, FIN-90014 Oulu, Finland\\}

\vspace*{0.8cm}
 
\mbox{\bf Abstract}

\end{centering}

\vspace*{0.3cm}
 
\noindent
We update Monte Carlo simulations of the three-dimensional 
SU(3) + adjoint Higgs theory, by extrapolating carefully to the infinite
volume and continuum limits, in order to estimate the contribution
of the infrared modes to the pressure of hot QCD. 
The sum of infrared contributions beyond the known 4-loop order 
turns out to be a smooth function, of a reasonable magnitude and
specific sign. Unfortunately, adding this function to the known 
4-loop terms does not improve the match to four-dimensional lattice 
data, in spite of the fact that  other quantities, such as 
correlation lengths, spatial string tension, or quark 
number susceptibilities, work well within the same setup.
We outline possible ways to reduce the mismatch.  
\vfill
\noindent
 
%
%
%
%

\vspace*{1cm}
 
\noindent
November 2008

\vfill

\end{titlepage}

%
\section{Introduction}
\la{se:introduction}

Motivated primarily by experimental heavy ion collision
programs at RHIC, LHC and FAIR, there are on-going  
large-scale efforts to determine the pressure of hot Quantum 
Chromodynamics (QCD) through numerical Monte Carlo simulations
of the full four-dimensional (4d) three-flavour~\cite{Nf3}--\cite{Nf3c} 
or even four-flavour~\cite{Nf4} theory.  
Due to the significant cost of simulating light 
dynamical fermions near the chiral limit, these efforts typically
make use of so-called staggered quarks. 
At any finite lattice spacing, the staggered action does 
not possess the flavour symmetries of the continuum theory, 
and is also problematic in the flavour-singlet sector. Although
these problems will eventually be overcome by the use of less
compromised fermion
discretizations, it would appear welcome in the meanwhile to explore 
complementary avenues as well, in order to offer timely 
crosschecks on the results that are being produced. 

One potentially useful avenue
in this respect is provided by effective field 
theory methods. In the limit of a high temperature, $T$, 
and a small gauge coupling, $g$, QCD develops a hierarchy 
of three momentum scales, $\pi T \gg gT \gg g^2 T/\pi$.
The largest scale, $\pi T$,  can be systematically integrated out, 
yielding a so-called Dimensionally Reduced effective field
theory~\cite{dr1,dr2}, or ``Electrostatic QCD'' (EQCD)~\cite{bn}.
EQCD has previously been used to determine non-perturbatively
quantities such as spatial correlation lengths~\cite{xis}, 
the spatial string tension~\cite{mt}--\cite{gE2}, 
and quark number susceptibilities~\cite{ahkr}. 
In all cases, a surprisingly good match to the results of 4d lattice 
simulations was found, even down to temperatures very close 
to that of the deconfining crossover, $\Tc$ 
(see, e.g., refs.~\cite{rbc,eh}).
We would therefore like to explore 
the extent to which a similar success could be achieved for 
the pressure (within pure Yang-Mills theory), and subsequently
perhaps apply the same methods to physical QCD. 

A way to apply EQCD to the study of the pressure on the non-perturbative
level, as well as first numerical results, were put forward a number
of years ago~\cite{a0cond}. The practical results of ref.~\cite{a0cond}
suffered from two problems, however: certain 4-loop logarithmic terms, 
whose form was not understood at the time, were effectively
missed~\cite{gsixg}; and the systematic error from the continuum 
extrapolation carried out was probably underestimated, given that 
the approach to the continuum limit has turned out to be more delicate
than anticipated~\cite{nspt_a02}. The purpose of this paper is, therefore, 
to update the analysis of ref.~\cite{a0cond}, both by making use of the new 
theoretical ingredients in refs.~\cite{gsixg,nspt_a02}, and by increasing
the numerical effort manifold. Unfortunately, the match to 4d lattice 
data does not improve despite all these efforts. On the other hand, 
the fact that systematic errors are now under control, implies that 
the discrepancy needs to be taken seriously, and this offers
us the possibility to speculate on the kind of physics that might 
be missing in our approach. In particular, we wish to discuss  
the role of higher dimensional operators within the EQCD framework, 
as well as the new qualitative features that should be expected from
more radically improved effective theories.  

We start by specifying the general setup of our approach
(\se\ref{se:setup}); go on to discuss the details of the lattice
formulation within EQCD (\se\ref{se:simulations}); present the 
main results (\se\ref{se:results}); and conclude with 
a discussion and outlook (\se\ref{se:conclusions}).

%
\section{General setup}
\la{se:setup}

At high temperatures, 
the pressure of QCD can be written as~\cite{bn}
\be
 p_\rmi{QCD}(T) = p_\rmi{hard}(T) 
                + p_\rmi{soft}(T) 
 \;. \la{pQCD}
\ee
Here $p_\rmi{hard}$ is a matching coefficient 
(defined in the $\msbar$ scheme) which gets contributions 
only from the hard scale, $k\sim \pi T$, and is computable in 
perturbation theory, while 
\be
 p_\rmi{soft}(T) 
  \equiv  \biggl\{ \lim_{V \to \infty} \frac{T}{V} 
 \ln \int \! {\cal D} A_i^a {\cal D} A_0^a  
 \, \exp\Bigl( -S_\rmi{E} 
 \Bigr) \biggr\}_\tinymsbar \;,
 \la{fMSdef}
\ee
where $V = \int\! {\rm d}^d\vec{x}$ is the $d$-dimensional volume
($d \equiv 3 - 2 \epsilon$), 
represents the contributions of the soft scales. The effective 
action can be written as 
\ba 
 S_\rmi{E} & = &  
 \int \! {\rm d}^d x\, \biggl\{
 \fr12 \tr [F_{ij}^2 ]+ \tr [D_i,A_0]^2 + 
 m_3^2\tr [A_0^2] 
 +\lambda_{3} (\tr [A_0^2])^2
 + ... \biggr\}
 \; . 
 \hspace*{0.5cm} \la{eqcd}
\ea
Here 
$F_{ij} = (i/g_{3}) [D_i,D_j]$, 
$D_i = \partial_i - i g_{3} A_i$, 
$A_i = A^a_i T^a$, 
$A_0 = A^a_0 T^a$, 
and $T^a$ are hermitean generators of SU(3). In the following 
we set $\epsilon\to 0$ in all finite quantities, whereafter
the dimensionalities of $g_3^2$ and $\lambda_3$ are GeV.

Now, $p_\rmi{soft}$ is scale-dependent, just like $p_\rmi{hard}$. 
Within the truncated form of \eq\nr{eqcd}, the scale dependence 
can be worked out explicitly~\cite{aminusb}. 
Defining the dimensionless ratios
\ba
 x  & \equiv & \frac{\lambda_3}{g_3^2}
 \;, \la{xdef} \\ 
 y & \equiv & \frac{m_3^2(\bmu=g_3^2)}{g_3^4}
 \;, \la{ydef} 
\ea
and ignoring terms of $\rmO(g^8)$ in terms of 4d power counting, 
we can write
\be
 p_\rmi{soft}(T) = 
 - T g_3^6 
 \biggl\{ 
   \mathcal{F}_\tinymsbar(x,y) + \frac{y\, d_A C_A}{(4\pi)^2}
   \ln \frac{\bmu}{g_3^2} - \frac{d_A C_A^3}{(4\pi)^4}  
  \biggl( 
  \frac{43}{3} - \frac{27}{32} \pi^2 \biggl) 
  \ln\frac{\bmu}{g_3^2}  
 \biggr\} 
 \;, \la{p_soft_splitup}
\ee
where $\bmu$ is the $\msbar$ scale parameter; 
$\mathcal{F}_\tinymsbar$ is by definition the 
vacuum energy density of the (truncated) EQCD, computed
with $\bmu = g_3^2$ and scaled dimensionless by dividing with $g_3^6$; 
and $d_A \equiv \Nc^2-1, C_A \equiv \Nc, \Nc \equiv 3$. 
The function $\mathcal{F}_\tinymsbar$ can, 
in turn, be written as 
\be
 \mathcal{F}_\tinymsbar(x,y)
 = 
 - \frac{d_A C_A^3}{(4\pi)^4}
 \biggl[ 
 \biggl( \frac{43}{12} - \frac{157}{768} \pi^2 \biggr) 
 \ln\frac{1}{2 C_A} + \bG
 \biggr] + 
 \mathcal{F}_\tinymsbar^\rmi{4-loop}(x,y) + 
 \mathcal{F}_\tinymsbar^{\mathcal{R}}(x,y) 
 \;. \la{F_splitup}
\ee
Here $\bG = -0.2\pm 0.8$ is 
the non-perturbative ``Linde term'' from 
the pure three-dimensional Yang-Mills theory~\cite{linde,gpy}, 
estimated numerically in refs.~\cite{plaq}--\cite{nspt_mass}, while 
$\mathcal{F}_\tinymsbar^\rmi{4-loop}$ is the 4-loop 
perturbative contribution sensitive to the adjoint 
Higgs field $A_0$~\cite{aminusb}, specified for completeness
in appendix A. The remainder, $\mathcal{F}_\tinymsbar^{\mathcal{R}}$, 
is what we address in the following.  

For dimensional reasons, the function 
$\mathcal{F}_\tinymsbar^{\mathcal{R}}$ necessarily 
depends on the parameter $y$ (it gets contributions from five loops
and beyond and, before the rescaling with $g_3^6$, therefore contains
terms of the form $g_3^{8+2 n}/m_3^{1+n}$, with $n\ge 0$). 
Thus no information is lost
(i.e.\ no integration constant is needed) 
if we take a partial derivative with respect to $y$, 
leading to  a condensate:
\be 
 \partial_{y} \mathcal{F}_\tinymsbar^{\mathcal{R}}(x,y)
  =
 \langle \tr[\hat A_0^2] 
 \rangle_\tinymsbar^{\mathcal{R}}
 \; \equiv \;
 \Bigl\langle \tr[\hat A_0^2]
  \Bigr\rangle_{\tinymsbar,\bmu=g_3^2}
 - 
  \Bigl\langle \tr[\hat A_0^2]
  \Bigr\rangle_{\tinymsbar,\bmu=g_3^2}^\rmi{4-loop}
 \;, \la{dy_F}
\ee
where $\hat A_0 \equiv {A_0}/{g_3}$. For a non-perturbative 
study, we need to change the scheme from $\msbar$ to the lattice, 
and then the relation in \eq\nr{dy_F} goes over into
\be
 \partial_{y} \mathcal{F}_\tinymsbar^{\mathcal{R}}(x,y)
 = 
 \lim_{a\to 0} 
 \Bigl\{ \Bigl\langle \tr [\hat A_0^2] \Bigr\rangle_a 
 - y^{\fr12} \, f_0(\hat m) - f_1(\hat m)  
 - y^{-\fr12} f_2(\hat m) - y^{-1} f_3(\hat m)
 \Bigr\} 
 \;, \la{fit2}
\ee
where
\be
 \amE \equiv 
 a g_3^2 y^{\fr12}
 \;, 
\ee
and $f_i$ are functions that have recently been 
determined numerically in ref.~\cite{nspt_a02}.
The task, therefore, is to measure 
$
 \langle \tr [\hat A_0^2] \rangle_a
$
on the lattice, for values of $x$ and $y$ corresponding to the 
physical finite-temperature QCD (cf.\ \fig\ref{fig:stripe}), 
and insert then the result in \eq\nr{fit2}. 

\begin{figure}[t]


\centerline{%
\epsfysize=8.0cm\epsfbox{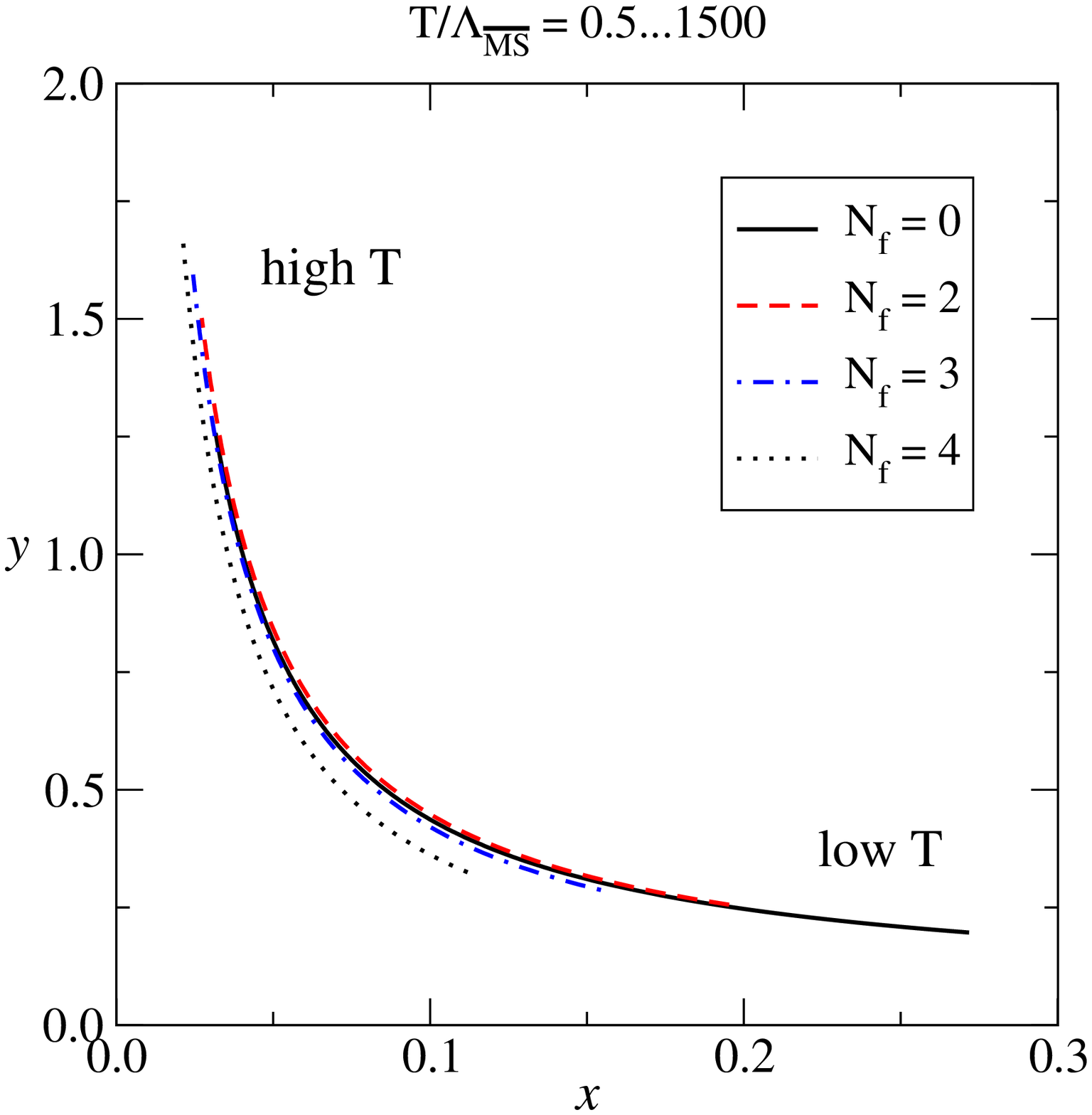}%
}


\caption[a]{\small
The stripes in the $(x,y)$-plane corresponding to 
various 4d theories in the temperature range
$T = (0.5 \ldots 1500) \Lambdamsbar$~\cite{adjoint}, 
with $\Nf$ denoting the number of massless quarks. 
The points in Table~1 
correspond to $\Nf = 0$ and $\Nf = 2$. 
The relation to the temperature is 
$
 y_\rmii{$(\Nf=0)$}\simeq 0.32\, [\log_{10}(T/\Lambdamsbar) + 0.91]
$,
$
 y_\rmii{$(\Nf=2)$}\simeq 0.38\, [\log_{10}(T/\Lambdamsbar) + 0.98]
$,
$
 y_\rmii{$(\Nf=3)$}\simeq 0.39\, [\log_{10}(T/\Lambdamsbar) + 1.04]
$.
}

\la{fig:stripe}
\end{figure}

%
\section{Lattice simulations}
\la{se:simulations}

The lattice study is carried out with the standard Wilson 
discretized action, 
\ba
 S_a \!\!& \equiv &\!\! 
 \beta \sum_\bfx\sum_{i<j}
 \biggl( 1-\fr13 \mathop{\mbox{Re}}\tr [P_{ij}(\bfx)]\biggl) 
 - {12\over\beta} \sum_{\bfx,i}\tr [ \hat A_0(\bfx)U_i(\bfx)\hat A_0(\bfx+i)
 U_i^\dagger(\bfx) ] 
 \nn \!\! & + & \!\! 
 \sum_\bfx \biggl\{ \alpha \tr [\hat A_0^2(\bfx)]+
 {216x_\rmi{latt} \over\beta^3}\Bigl(  \tr [\hat A_0^2(\bfx)]\Bigr)^2
 \biggr\} \;, 
 \la{lattaction}
\ea
where
$U_i({\bf x})$ is a link matrix; 
${\bf x}+i\equiv  {\bf x}+a\hat\e_i$, with $\hat\e_i$ 
a unit vector; $P_{ij}({\bf x})$ is the plaquette; and
\be
 \beta \equiv \frac{6}{g_3^2 a}
 \;. \la{beta}
\ee
The bare mass parameter is given by~\cite{contlatt}
\ba
 \alpha
 & = & {36\over\beta}
 \biggl\{ 1+{6\over\beta^2}y_\rmi{latt}
 -(6+10x_\rmi{latt}){3.175911535625\over
 4\pi\beta} 
 \nn & &{}
 -{3\over8\pi^2\beta^2}\Bigl[(60x_\rmi{latt}
 -20x_\rmi{latt}^2)(\ln\beta+0.08849)+
 34.768x_\rmi{latt}+36.130\Bigr]\biggr\}
 \;. \la{2lm2}
\ea
For historical reasons, we have implemented in most 
of the runs a partial $\rmO(a)$
improvement~\cite{moore_a}, by taking 
\be
 x_\rmi{latt}  \equiv   x+\fr{0.328432-0.835282x+1.167759x^2}\beta \;,
\ee
but for the mass parameter no improvement was carried out, since
the additive $\rmO(a)$ terms have not been determined: 
$
 y_\rmi{latt} \equiv y 
$.
The improvement of $x_\rmi{latt}$ turns out to play little practical role 
if properly taken into account in the subtraction of \eq\nr{fit2}; 
in fact we did not implement it in the last sets of runs, 
corresponding to $\Nf=2$ as well as to $\beta=240$ with $\Nf = 0$. 
In any case our setup conforms with ref.~\cite{nspt_a02}, 
where the functions $f_i$ in \eq\nr{fit2} were determined 
numerically for general $x_\rmi{latt}$.

The input parameters in any run are $x,y,\beta$.
The phase diagram of the system in this space has a ``disordered'', 
or symmetric phase, as well as a  symmetry 
broken phase~\cite{su3adj}. The simulations we carry 
out represent the physical QCD only in the symmetric
phase~\cite{adjoint}. It turns out that on the physical
stripe (\fig\ref{fig:stripe}) the symmetric phase is actually 
metastable, but strongly enough so for any practical effects
to be miniscule in large enough volumes 
(see, however, the discussion in \se\ref{se:conclusions}).

The three-dimensional SU(3) + adjoint Higgs theory is a confining 
gauge theory, and possesses a mass gap (for a practical 
demonstration see, e.g., ref.~\cite{xis}). Thereby all finite
volume effects must be exponentially suppressed. 
For completeness we have checked this explicitly: 
the condensate 
$
 \langle \tr [\hat A_0^2] \rangle_a
$
as a function of $\beta/N = 6/ g_3^2 L$, where $L$ is the extent
of the box, is shown in \fig\ref{fig:vol}.
No finite-volume effects are visible in 
the range $\beta/N < 1$, or $L > 6/g_3^2$, 
to which we restrict in the following. 
The parameters used for the production runs
are listed in Table~\ref{table:stats}.

\begin{figure}[t]


\centerline{%
\epsfysize=7.0cm\epsfbox{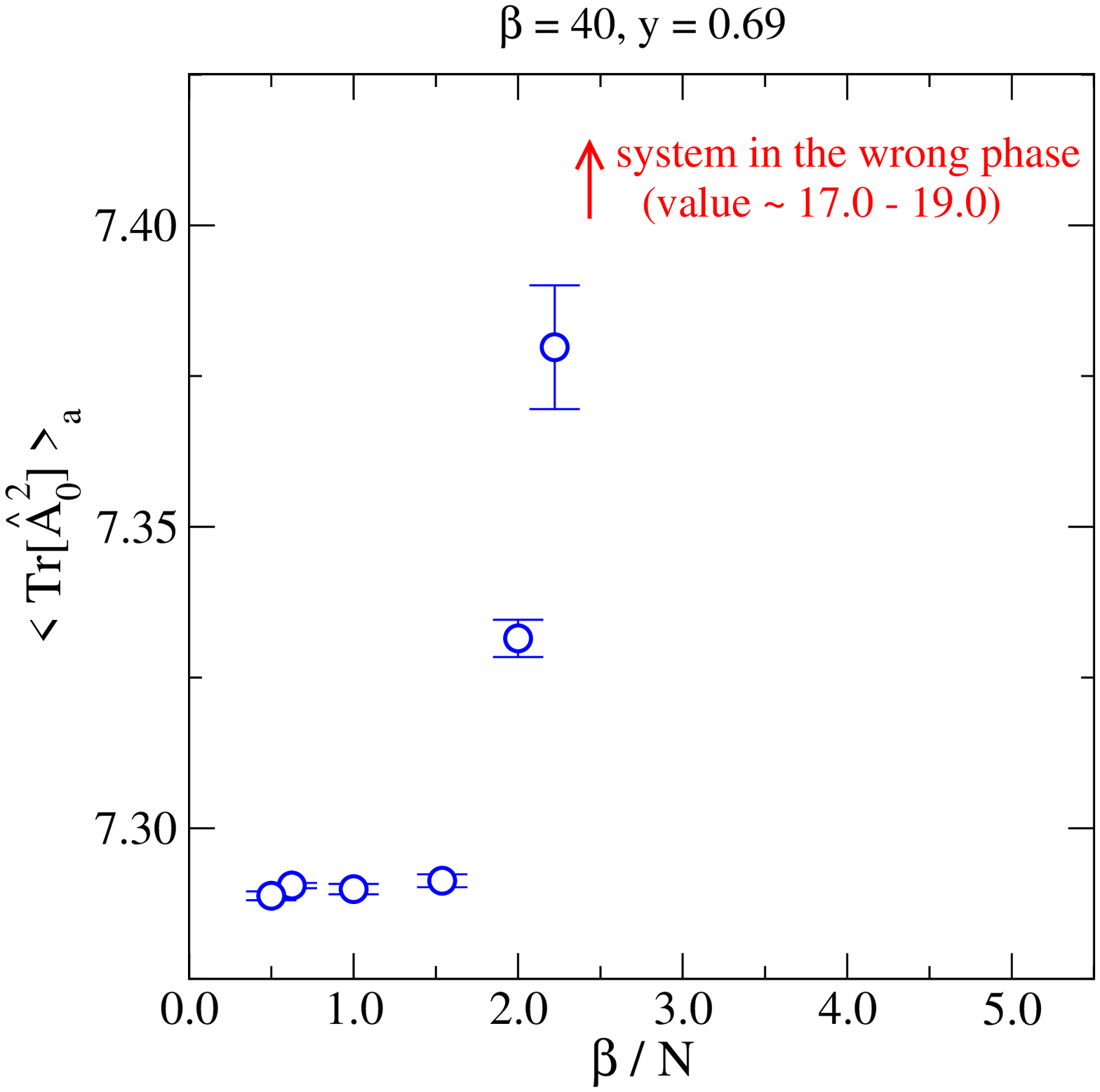}%
~~\epsfysize=7.0cm\epsfbox{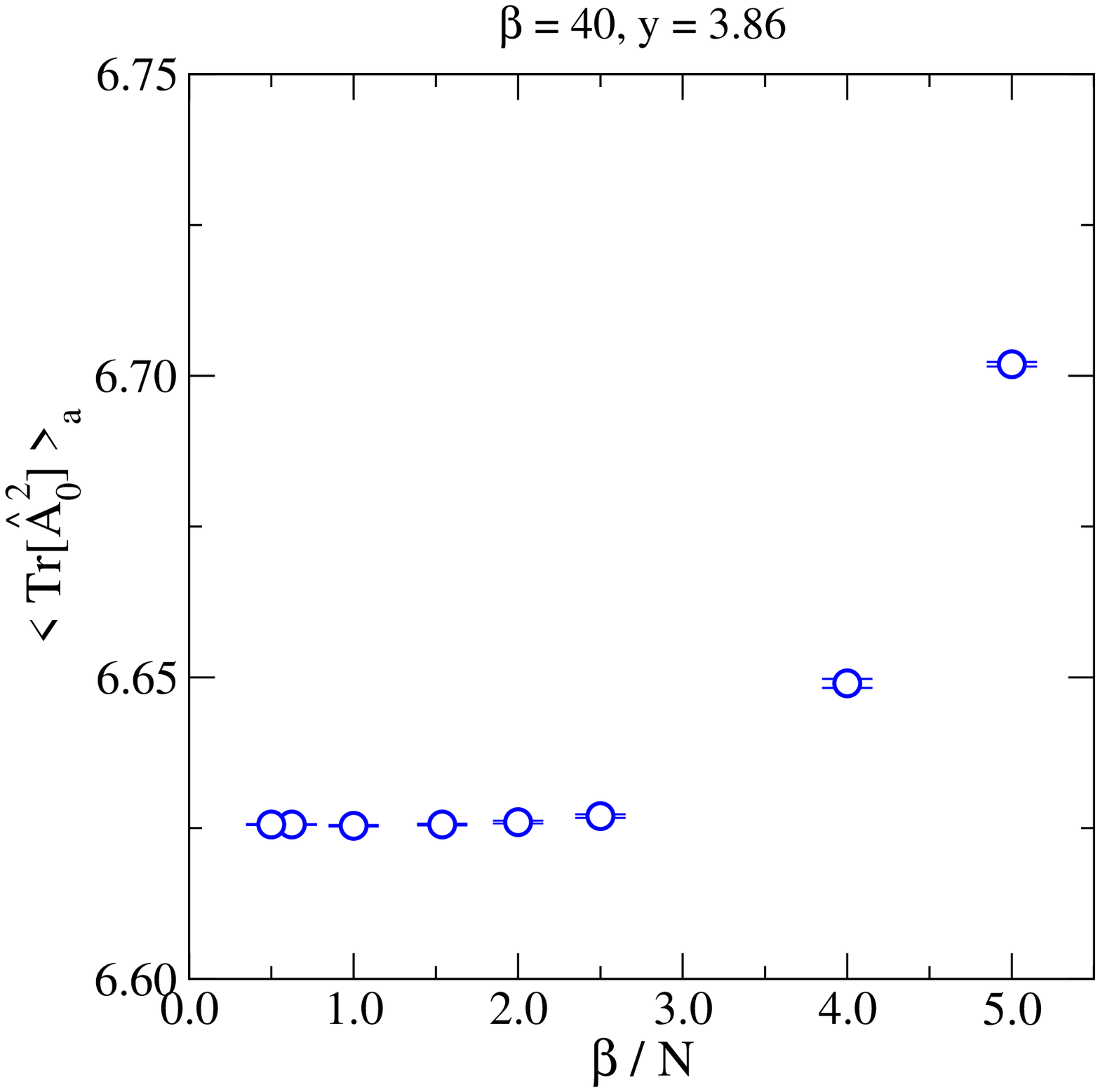}%
}


\caption[a]{\small
Finite-volume values for   
$
 \langle \tr [\hat A_0^2] \rangle_a
$,
as a function of the physical
extent $\beta/N = 6/ g_3^2 L$ of the box ($L=aN$), 
for a small $y$ (left) and large $y$ (right). No volume 
dependence is visible for $\beta/N < 1.0$. At small $y$ and 
small volumes, the metastability of the physical phase is too 
weak to hold the system there for any length of Monte Carlo time. 
}

\la{fig:vol}
\end{figure}

%
\begin{table}[!ht]

\small 

\begin{center}
\begin{tabular}{llllllllll} \hline  \\[-2mm]
 $x$ & $y$ &  $\beta_\rmii{$N$}$ \\[2mm] \hline \\[-1mm]
      0.0060 &       6.389567 &     
             24$_\rmii{48}\!\!$ &
             32$_\rmii{40, 64, 96}\!\!$ & 
             40$_\rmii{64, 80}\!\!$ &
             54$_\rmii{64, 96}\!\!$ &
             67$_\rmii{120}\!\!$ &
             80$_\rmii{96, 144}\!\!$ &
            120$_\rmii{144$^2$, 200}\!\!$ &
            240$_\rmii{512}$ \\
             &       $\hspace*{-0.15cm}^*$6.621393 &     
              &
             32$_\rmii{144}\!\!$ & 
             40$_\rmii{144}\!\!$ &
             54$_\rmii{144}\!\!$ &
             67$_\rmii{144}\!\!$ &
             80$_\rmii{144}\!\!$ &
            120$_\rmii{144, 256}\!\!$ &
              \\
      0.0075 &       5.123052 &
             24$_\rmii{48}\!\!$ &
             32$_\rmii{64}\!\!$ &
             40$_\rmii{80}\!\!$ &
             54$_\rmii{96}\!\!$ &
             67$_\rmii{120}\!\!$ &
             80$_\rmii{144}\!\!$ &
            120$_\rmii{144, 200}\!\!$ &
            240$_\rmii{512}$ \\
           &       $\hspace*{-0.15cm}^*$5.307970 &
              &
             32$_\rmii{144}\!\!$ &
             40$_\rmii{144}\!\!$ &
             54$_\rmii{144}\!\!$ &
             67$_\rmii{144}\!\!$ &
             80$_\rmii{144}\!\!$ &
            120$_\rmii{144, 256}\!\!$ &
             \\
      0.0100 &       3.856538 &
             24$_\rmii{48}\!\!$ &
             32$_\rmii{64}\!\!$ &
             40$_\rmii{64, 80}\!\!$ &
             54$_\rmii{96}\!\!$ &
             67$_\rmii{120}\!\!$ &
             80$_\rmii{144}\!\!$ &
            120$_\rmii{144, 200}\!\!$ &
            240$_\rmii{512}$ \\
         &       $\hspace*{-0.15cm}^*$3.994547 &
              &
             32$_\rmii{144}\!\!$ &
             40$_\rmii{144}\!\!$ &
             54$_\rmii{144}\!\!$ &
             67$_\rmii{144}\!\!$ &
             80$_\rmii{144}\!\!$ &
            120$_\rmii{144, 256}\!\!$ &
              \\
      0.0130 &       2.979720 &
             24$_\rmii{48}\!\!$ &
             32$_\rmii{40, 64}\!\!$ &
             40$_\rmii{80}\!\!$ &
             54$_\rmii{96}\!\!$ &
             67$_\rmii{120}\!\!$ &
             80$_\rmii{144$^2$}\!\!$ &
            120$_\rmii{144, 200}\!\!$ &
            240$_\rmii{512}$ \\
          &       $\hspace*{-0.15cm}^*$3.085255 &
              &
             32$_\rmii{144}\!\!$ &
             40$_\rmii{144}\!\!$ &
             54$_\rmii{144}\!\!$ &
             67$_\rmii{144}\!\!$ &
             80$_\rmii{144}\!\!$ &
            120$_\rmii{144, 256}\!\!$ &
             \\
      0.0200 &       1.956765 &
             24$_\rmii{48}\!\!$ &
             32$_\rmii{40, 64}\!\!$ &
             40$_\rmii{80}\!\!$ &
             54$_\rmii{96}\!\!$ &
             67$_\rmii{120}\!\!$ &
             80$_\rmii{144}\!\!$ &
            120$_\rmii{144, 200}\!\!$ &
            240$_\rmii{512}$ \\
        &       $\hspace*{-0.15cm}^*$2.024413 &
              &
             32$_\rmii{144}\!\!$ &
             40$_\rmii{144}\!\!$ &
             54$_\rmii{144}\!\!$ &
             67$_\rmii{144}\!\!$ &
             80$_\rmii{144}\!\!$ &
            120$_\rmii{144, 256}\!\!$ &
              \\
      0.0260 &       1.518356 &
             24$_\rmii{48}\!\!$ &
             32$_\rmii{64, 120}\!\!$ &
             40$_\rmii{80}\!\!$ &
             54$_\rmii{108}\!\!$ &
             67$_\rmii{120}\!\!$ &
             80$_\rmii{120}\!\!$ &
            120$_\rmii{200}\!\!$ &
            240$_\rmii{512}$ \\
      0.0300 &       1.323508 &
             24$_\rmii{48}\!\!$ &
             32$_\rmii{64, 120}\!\!$ &
             40$_\rmii{80}\!\!$ &
             54$_\rmii{108}\!\!$ &
             67$_\rmii{120}\!\!$ &
             80$_\rmii{120}\!\!$ &
            120$_\rmii{200}\!\!$ &
            240$_\rmii{512}$ \\
      0.0350 &       1.142577 &
             24$_\rmii{48}\!\!$ &
             32$_\rmii{40, 64, 96}\!\!$ &
             40$_\rmii{64, 80}\!\!$ &
             54$_\rmii{84, 96}\!\!$ &
             67$_\rmii{120}\!\!$ &
             80$_\rmii{84, 144}\!\!$ &
            120$_\rmii{144, 200}\!\!$ &
            240$_\rmii{512}$ \\
        &       $\hspace*{-0.15cm}^*$1.180070 &
               &
             32$_\rmii{144}\!\!$ &
             40$_\rmii{144}\!\!$ &
             54$_\rmii{144}\!\!$ &
             67$_\rmii{144}\!\!$ &
             80$_\rmii{144}\!\!$ &
            120$_\rmii{144, 256}\!\!$ &
             \\
      0.0450 &       0.901336 &
             24$_\rmii{48}\!\!$ &
             32$_\rmii{64, 120}\!\!$ &
             40$_\rmii{80}\!\!$ &
             54$_\rmii{108}\!\!$ &
             67$_\rmii{120}\!\!$ &
             80$_\rmii{120}\!\!$ &
            120$_\rmii{200}\!\!$ &
            240$_\rmii{512}$ \\
      0.0600 &       0.690251 &
             24$_\rmii{48}\!\!$ &
             32$_\rmii{64}\!\!$ &
             40$_\rmii{64, 80}\!\!$ &
             54$_\rmii{96}\!\!$ &
             67$_\rmii{120}\!\!$ &
             80$_\rmii{144}\!\!$ &
            120$_\rmii{200}\!\!$ &
            240$_\rmii{512}$ \\
          &       $\hspace*{-0.15cm}^*$0.710991 &
               &
             32$_\rmii{144}\!\!$ &
             40$_\rmii{144}\!\!$ &
             54$_\rmii{144}\!\!$ &
             67$_\rmii{144}\!\!$ &
             80$_\rmii{144}\!\!$ &
            120$_\rmii{144, 256}\!\!$ &
              \\
      0.0860 &       0.498801 &
             24$_\rmii{48}\!\!$ &
             32$_\rmii{64, 120}\!\!$ &
             40$_\rmii{80}\!\!$ &
             54$_\rmii{108}\!\!$ &
             67$_\rmii{120}\!\!$ &
             80$_\rmii{120}\!\!$ &
            120$_\rmii{200}\!\!$ &
            240$_\rmii{512}$ \\
      0.1000 &       0.436948 &
             24$_\rmii{48}\!\!$ &
             32$_\rmii{64}\!\!$ &
             40$_\rmii{80}\!\!$ &
             54$_\rmii{96}\!\!$ &
             67$_\rmii{120}\!\!$ &
             80$_\rmii{144}\!\!$ &
            120$_\rmii{200}\!\!$ &
            240$_\rmii{512}$ \\
          &       $\hspace*{-0.15cm}^*$0.448306 &
               &
             32$_\rmii{144}\!\!$ &
             40$_\rmii{144}\!\!$ &
             54$_\rmii{144}\!\!$ &
             67$_\rmii{144}\!\!$ &
             80$_\rmii{144}\!\!$ &
            120$_\rmii{144, 256}\!\!$ &
               \\
      0.1300 &       $\hspace*{-0.15cm}^*$0.357377 &
              &
             32$_\rmii{176}\!\!$ &
             40$_\rmii{176}\!\!$ &
             54$_\rmii{176}\!\!$ &
             67$_\rmii{176}\!\!$ &
             80$_\rmii{176, 320}\!\!$ &
            120$_\rmii{176, 320}\!\!$ &
  \\[2mm] \hline
\end{tabular}
\end{center}

\normalsize

\noindent
\caption[a]{\small
The continuum parameters $x,y$ (cf.\ \eqs\nr{xdef}, \nr{ydef}); 
the lattice couplings $\beta$ (cf.\ \eq\nr{beta}); 
and the box sizes $N$ ($V = a^3 N^3$) used for the production runs. 
Values marked with a star correspond to $\Nf = 2$, others to $\Nf = 0$.
In the few cases where the box size
has the superscript 2, two independent runs were launched.  
In total, our sample consists of 186 lattices. 
}

\la{table:stats}
\end{table}
%

%
\section{Numerical results}
\la{se:results}

\begin{figure}[t]


\centerline{%
\epsfysize=7.0cm\epsfbox{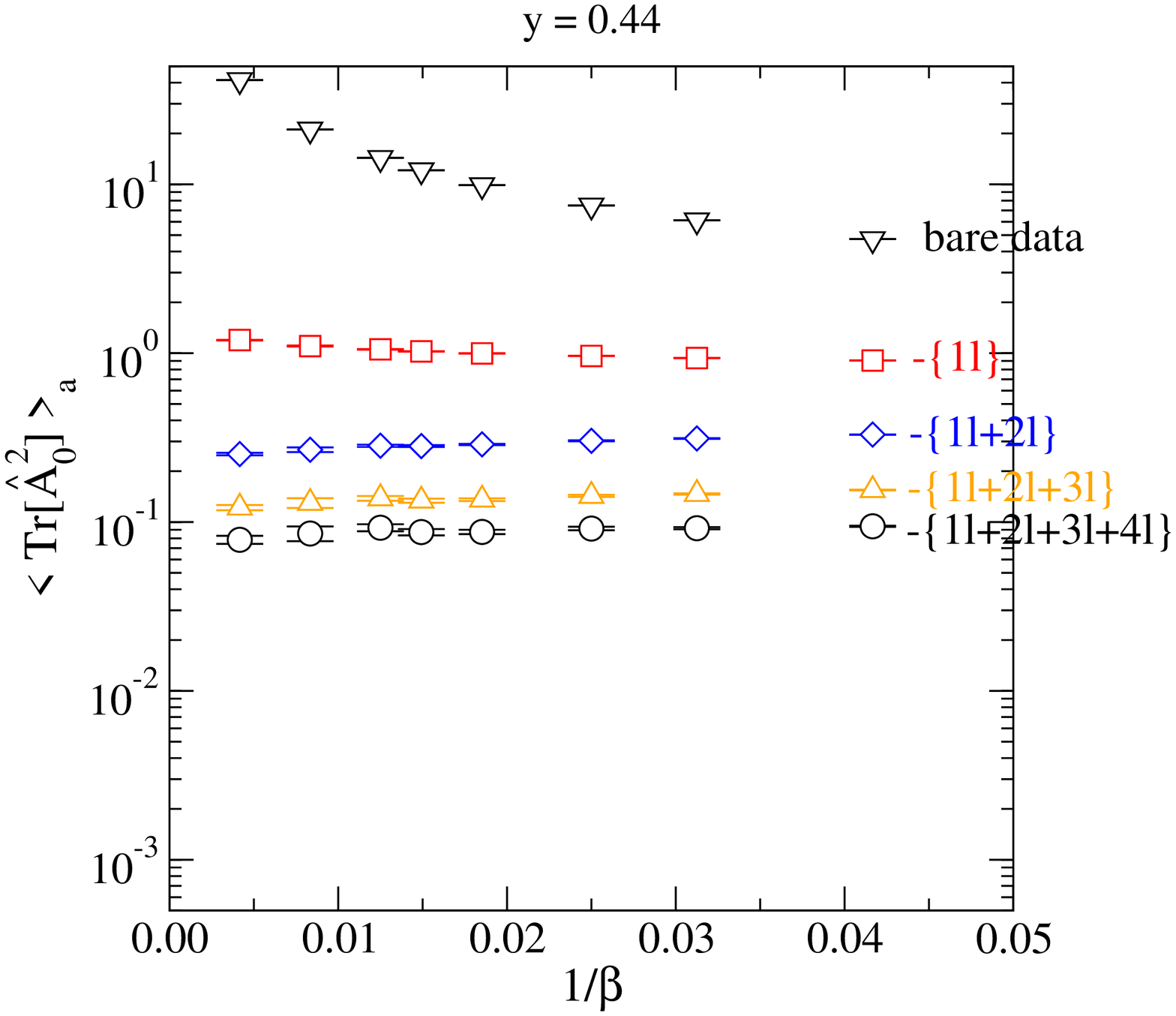}%
~~\epsfysize=7.0cm\epsfbox{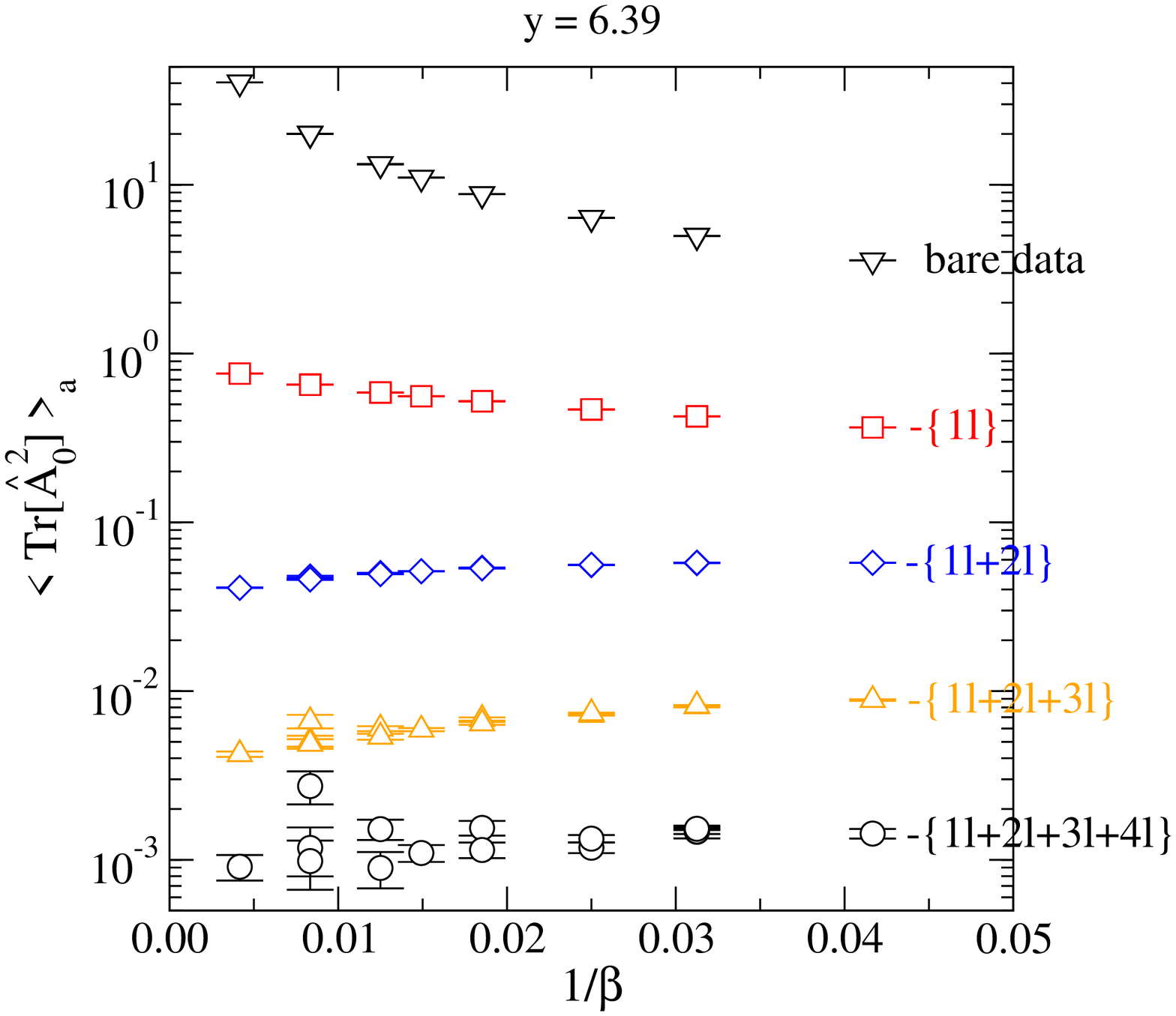}%
}

\vspace*{0.5cm}


\caption[a]{\small
Shown is $\langle\tr[\hat A_0^2]\rangle_a$ 
at a small and large $y$, after 
the consecutive subtraction of the functions 
$f_0,f_1,f_2,f_3$ of \eq\nr{fit2}, corresponding
to 1-loop, 2-loop, 3-loop and 4-loop effects, respectively. }

\la{fig:subtractions}
\end{figure}

In \fig\ref{fig:subtractions} we show
$\langle\tr[\hat A_0^2]\rangle_a$ at two values of $y$, 
as a function of the lattice spacing $1/\beta = a g_3^2/6$, 
and after the subtraction of the various terms in \eq\nr{fit2}.  
The figure indicates that after the subtractions, a continuum
limit ($1/\beta\to 0$) can be taken; and that perturbation 
theory does converge even at the smallest $y$ (corresponding
to the lowest temperature), in the sense that each subtraction
is smaller than the previous one, and that the remainder is 
at most of the same order as the last subtraction. Increasing $y$, 
perturbation theory converges faster, and the significance 
loss due to the subtractions becomes substantial.

\begin{figure}[t]


\centerline{%
\epsfysize=8.0cm\epsfbox{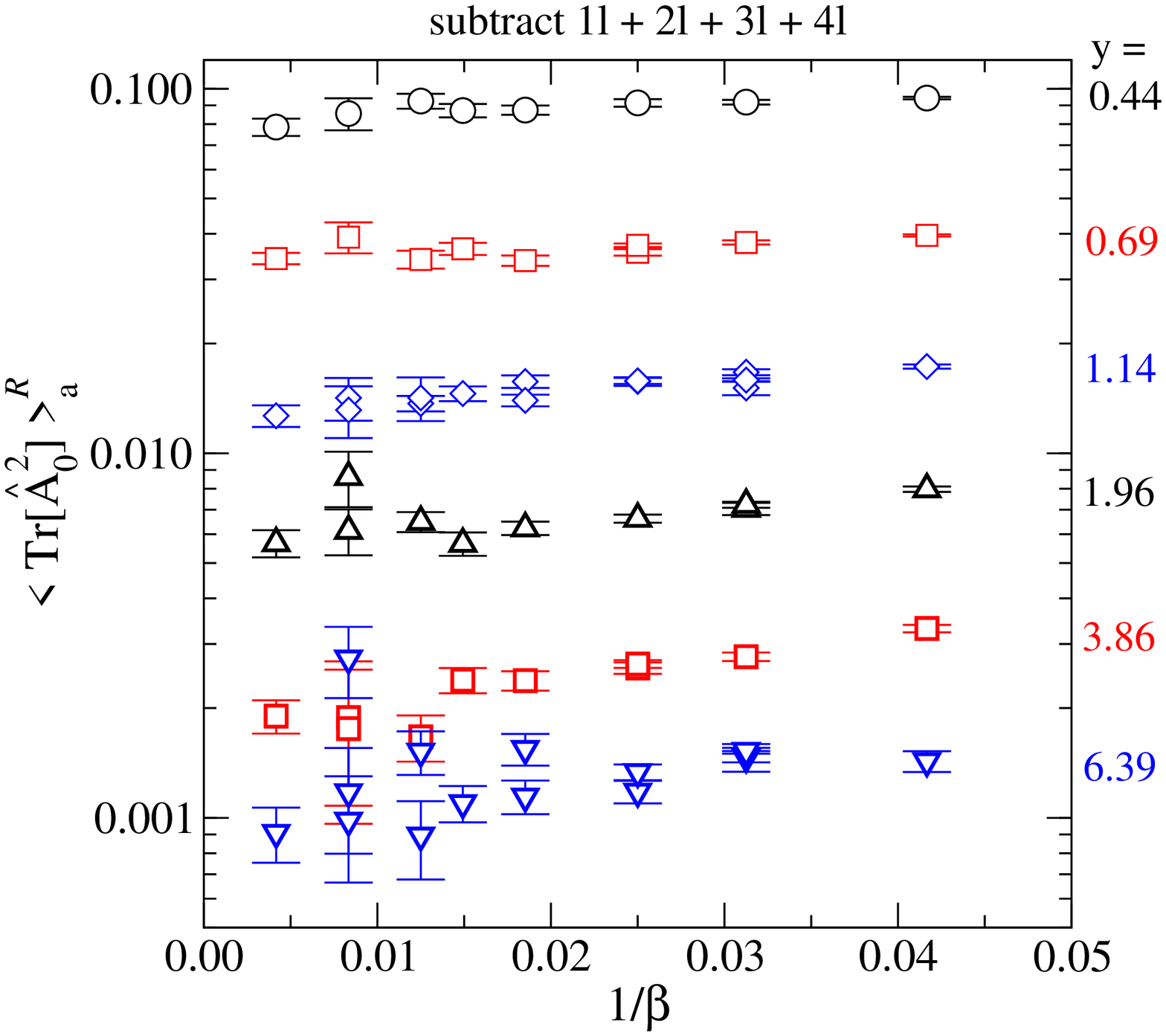}%
}


\caption[a]{\small
A magnification of the fully subtracted
results from data of the type in \fig\ref{fig:subtractions}, 
for selected values of $y$, indicated on the right.
Multiple data points at the same $\beta$
correspond to independent runs, or different volumes
(cf.\ Table~\ref{table:stats}). 
}

\la{fig:remainder}
\end{figure}

In \fig\ref{fig:remainder} we show a magnification of the fully 
subtracted results, for selected values of $y$. This highlights 
the regular-looking approach to the continuum limit. 

\begin{figure}[t]


\centerline{%
\epsfysize=8.0cm\epsfbox{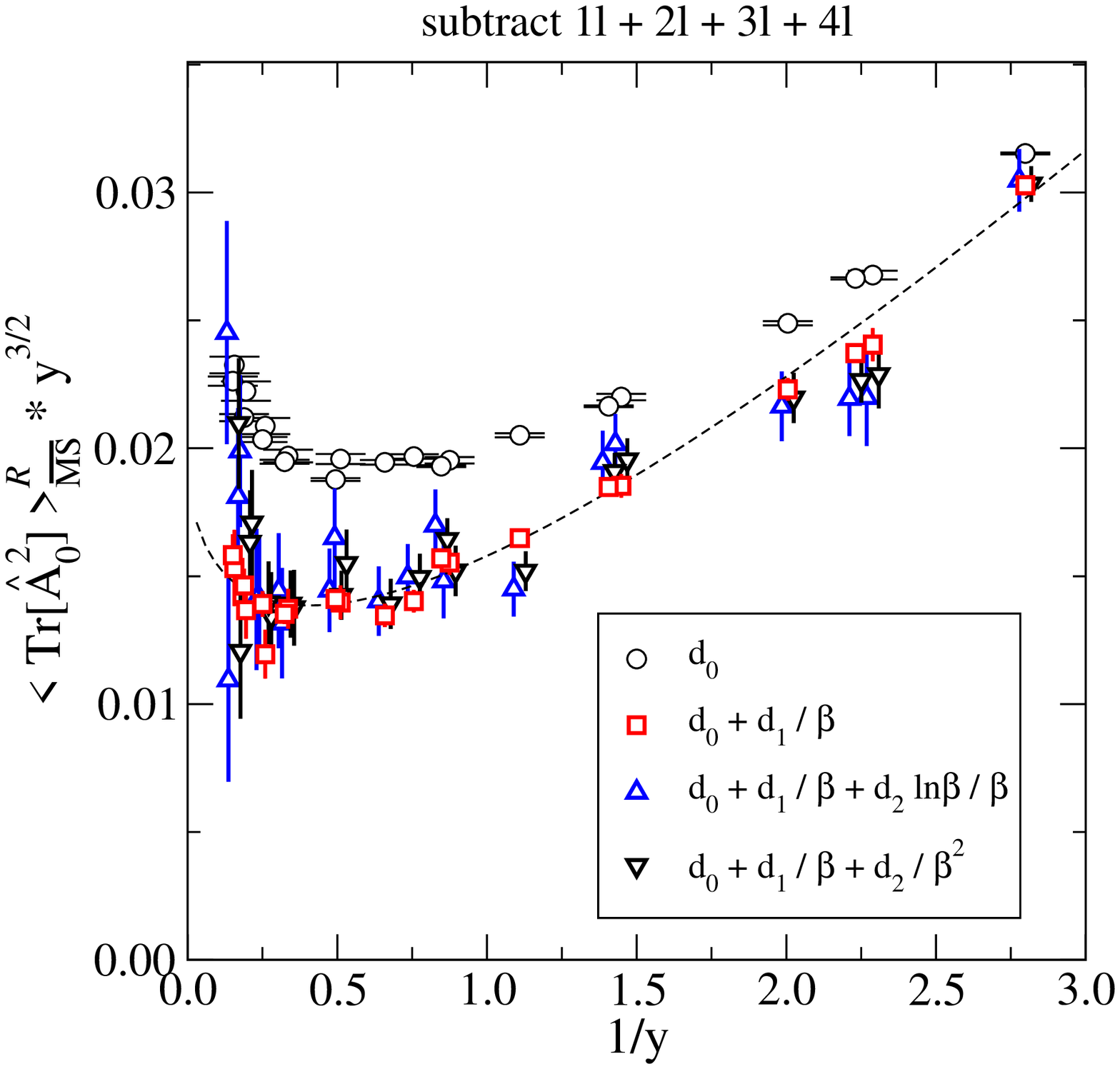}%
}


\caption[a]{\small
Continuum ($1/\beta = 0$) intercepts for data of the type in 
\fig\ref{fig:remainder}, multiplied by $y^{3/2}$, obtained from 
fits with different $\beta$-dependences, as indicated in the legend. 
Some points have been slightly displaced for better visibility.
The dashed curve is from \eq\nr{A0R_fit}. }

\la{fig:intercept}
\end{figure}

In order to carry out the continuum extrapolation, we have tested
four different fit functions: a constant fit,  a fit linear in $1/\beta$, 
a fit including a linear term and the logarithm $(\ln\beta )/ \beta$, 
as well as a quadratic fit. The results are shown in \fig\ref{fig:intercept}.
We observe that all dependences apart
from the first one lead to consistent results; the situation is thus quite 
different from ref.~\cite{ahkr}, where fits allowing for a logarithmic
term were the only ones leading to a sensible outcome. In other words, 
the 3-loop and 4-loop subtractions in \eq\nr{fit2}, which were not
available at the time of ref.~\cite{ahkr}, have effectively removed
any possible logarithmic terms. In the following, intercepts obtained from 
linear extrapolations (squares) will be used. The dashed line
in \fig\ref{fig:intercept} shows the curve
\be
 \langle \tr[\hat A_0^2] 
 \rangle_\tinymsbar^{\mathcal{R}}  \simeq \frac{1}{y^{3/2}}
 \biggl(
  c_1 + \frac{ c_2 }{ y^{1/2} } + \frac{ c_3 }{ y }
 \biggr)
 \;, \la{A0R_fit} 
\ee 
with 
$c_1 = 0.0200(8)$,  $c_2 = -0.0191(14)$,  $c_3 = 0.0149(6)$.
This representation describes our data well 
($\chi^2/\mbox{d.o.f.} = 2.6$) in the whole $y$-range considered.

The next task is to integrate \eq\nr{fit2} in order to determine
the function $\mathcal{F}_\tinymsbar^{\mathcal{R}}$. Since \eq\nr{fit2}
contains a partial rather than a total derivative, and on the physical 
stripe the parameter $x$ changes (cf.\ \fig\ref{fig:stripe}), 
an integration is 
strictly speaking not possible\footnote{%
 In ref.~\cite{a0cond} this problem was
 tackled by also measuring the condensate 
 $\partial_x \mathcal{F}_\tinymsbar = 
 \langle (\tr[\hat A_0^2])^2 
 \rangle_a - \;...\;$. However, some of the renormalization
 constants needed for the $\msbar$ conversion remain unknown, 
 and in any case the practical effects from this condensate 
 appeared to be too small to change the qualitative behaviour. 
 };  
however, the leading $x$-dependent term 
of  $\mathcal{F}_\tinymsbar^{\mathcal{R}}$
corresponds to a contribution  
of the type $p_\rmi{soft} \sim g_3^6 \lambda_3/m_3 \sim g^9 T^4$, 
which is of higher order than other effects we have ignored. More 
importantly, all effects containing the parameter $x$ are numerically
subdominant for $\Nc = 3$ (cf.\ \eq\nr{calFxy}).
Thereby, close to the physical
stripe, the remainder in 
\eq\nr{A0R_fit} can simply be integrated to  
\be
 \mathcal{F}_\tinymsbar^{\mathcal{R}}(x,y)  
 \simeq 
 -\frac{2}{y^{1/2} } \biggl( 
 {c_1} +  \frac{c_2 }{ 2 y^{1/2} } +  \frac{c_3 }{ 3 y }
 \biggr) 
 \;. \la{calFR}
\ee
We note that the numerical value of 
the integral $\mathcal{F}_\tinymsbar^{\mathcal{R}}$ is only weakly
dependent on the ansatz used for the continuum extrapolation.

We can now compare our result with 4d lattice data. 
In the following we consider pure glue~\cite{Nf0}, for which 
systematic errors are best under control. To get the EQCD result,
we need to evaluate both $p_\rmi{hard}$ and $p_\rmi{soft}$
(\eq\nr{pQCD}). In practice, we take part of the terms in $p_\rmi{soft}$, 
namely those specified explicitly in \eqs\nr{p_soft_splitup}, 
\nr{F_splitup}, and combine them with 
the hard contribution $p_\rmi{hard}$, specified in \eqs(5), (6) of 
ref.~\cite{pheneos}. Given that the non-perturbative constant
$\bG$ contains numerical errors and that $p_\rmi{hard}$
contains a perturbative constant, denoted by $\Delta_\rmi{hard}$ in 
ref.~\cite{pheneos}, which remains unknown, we treat the combination
$
 \Delta_\rmi{hard} + 
 d_A C_A^3 
 \bG
$
as a free parameter. 
The remaining soft contribution, 
$
 \delta p_\rmi{soft} = - T g_3^6 \mathcal{F}_\tinymsbar 
$
(cf.\ \eq\nr{p_soft_splitup}), 
is given by the sum of \eqs\nr{calFR}, \nr{calFxy}. 
The free parameter 
$
 \Delta_\rmi{hard} + 
 d_A C_A^3 
 \bG
$
is fixed by minimizing the $\chi^2$-difference
of 4d lattice data~\cite{Nf0} and our full 
result in the range $T/\Tc \ge 3.0$.
The outcome of the fit is shown in \fig\ref{fig:pheneos}. 
(It corresponds to 
$
 \Delta_\rmi{hard} + 
 d_A C_A^3 
 \bG = -7.24.
$) 

\begin{figure}[t]


\centerline{%
\epsfysize=7.0cm\epsfbox{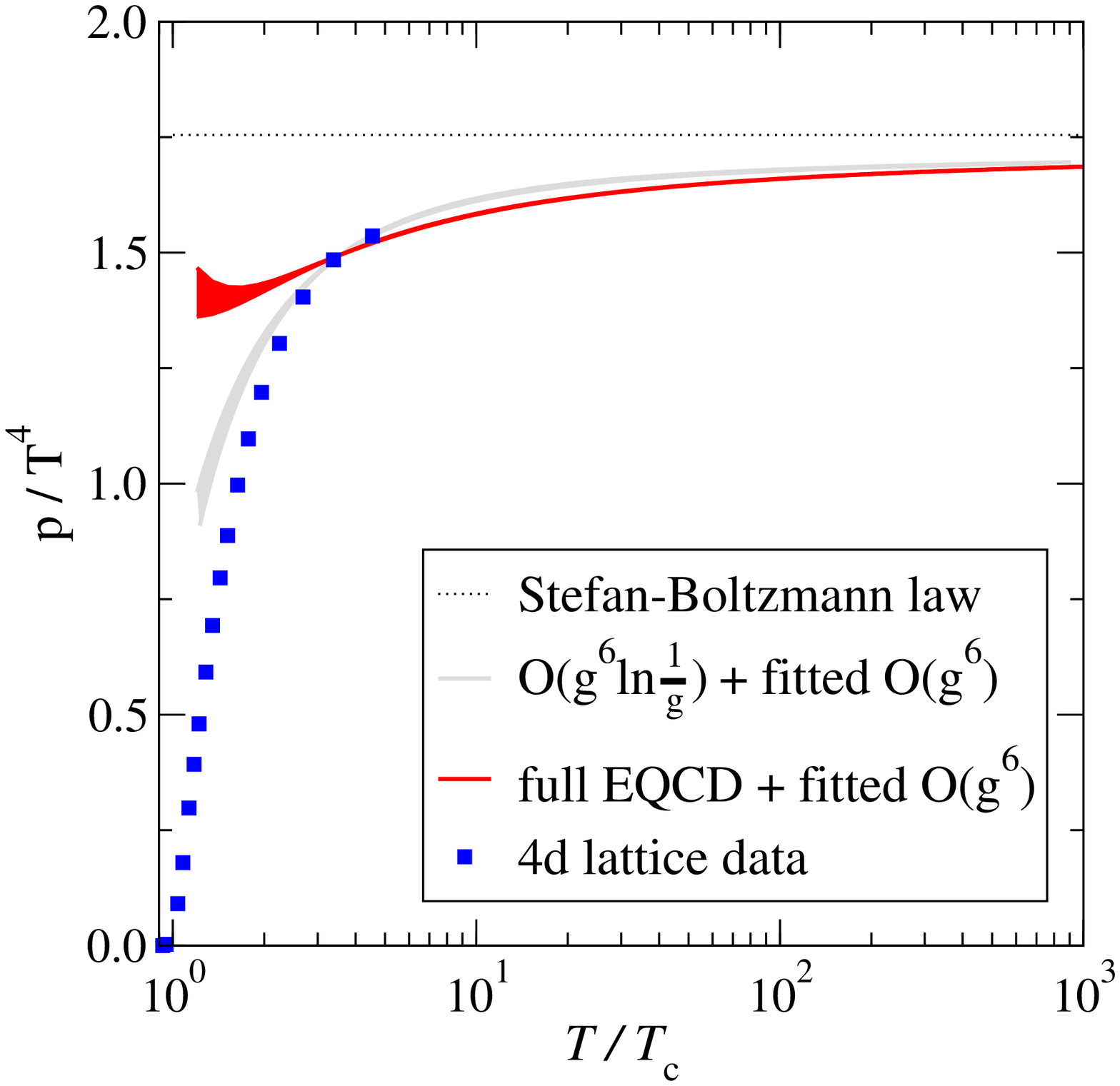}%
~~~~~\epsfysize=7.0cm\epsfbox{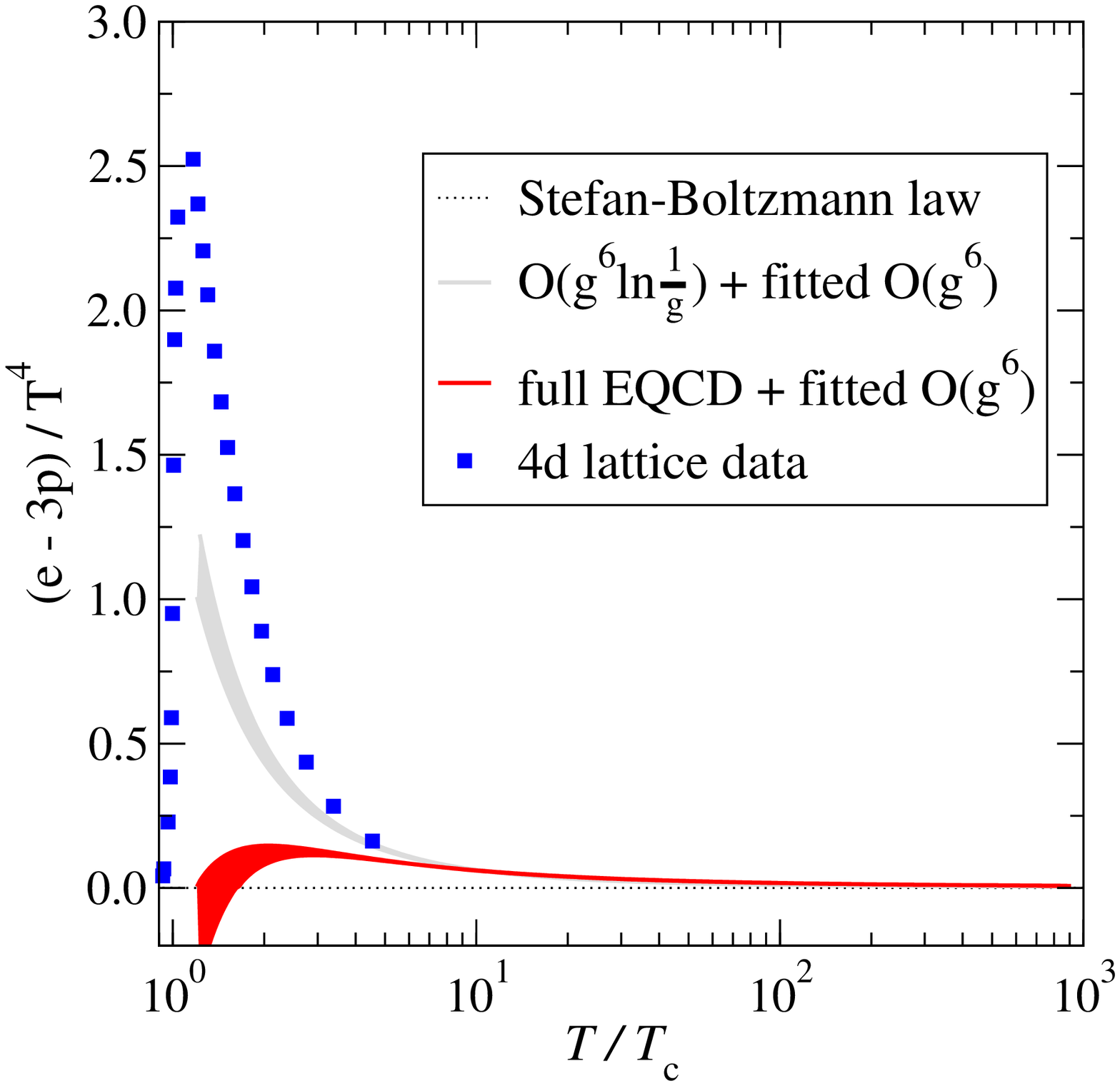}%
}


\caption[a]{\small
Left: the best fit of our result to 4d lattice data 
for $p/T^4$ at $\Nf = 0$~\cite{Nf0}. 
Our result comes out as a function of $T/\Lambdamsbar$; for converting
to $T/\Tc$ we scanned the interval 
$\Tc/\Lambdamsbar = 1.10 \ldots 1.35$ (dark band).
For comparison, with the light band
we show the outcome for 
$
  \mathcal{F}_\tinymsbar^{\mathcal{R}} \equiv 0
$~\cite{pheneos}.
Right: the corresponding trace anomaly, 
$(e-3p)/T^4 = T {\rm d}(p/T^4)/{\rm d}T$. 
}

\la{fig:pheneos}
\end{figure}

It can be observed from \fig\ref{fig:pheneos} 
that our high-temperature result 
and the low-temperature 4d lattice result depart already
at $T\approx 3.2\Tc$, and that the fit 
is in general not particularly good. In fact, 
compared with the fit in ref.~\cite{pheneos} which assumed 
$\mathcal{F}_\tinymsbar^{\mathcal{R}} \equiv 0$, 
$\chi^2$ increases about 30-fold.
(The fit of ref.~\cite{pheneos} corresponds to 
$
 \Delta_\rmi{hard} + 
 d_A C_A^3 
 \bG = -1.84
$.)
The reason for the increase is that 
$\langle \tr[\hat A_0^2] 
 \rangle_\tinymsbar^{\mathcal{R}}$ 
in \eq\nr{A0R_fit} is positive, 
whereby $\mathcal{F}_\tinymsbar^{\mathcal{R}}$
in \eq\nr{calFR} is negative, whereby
$p_\rmi{soft}/T^4$ gets a contribution {\em increasing} rapidly
at small temperatures (cf.\ \eq\nr{p_soft_splitup}). 
This is completely unlike the behaviour of the 4d lattice data.

Now, one possible reason for the mismatch could be that, 
starting at $\rmO(g^7)$, higher order 
operators should be added to the EQCD action~\cite{gsixg}. It seems, 
however, that in practice such operators cannot change the result 
in a substantial way. Indeed, for $\Nf = 0$, the higher order 
operators were determined in ref.~\cite{sc}, and the only one
contributing at $\rmO(g^7 T^4)$ reads
\be
 \delta S_\rmi{E} \sim  \int \! {\rm d}^d x\,  
 \frac{g^2 C_A}{(4\pi)^4 T^2}
 \mathcal{D}_i^2 A^a_0 \, 
 \mathcal{D}_j^2 A^a_0
 \;. \la{Og7} 
\ee
Treating this perturbatively (as a ``vertex''), and working
in dimensional regularization, yields a contribution
\be
 \delta p_\rmi{soft} \sim 
 \frac{d_A C_A}{(4\pi)^4 T} \frac{g^2 m_3^5}{4\pi}
 \sim \frac{1}{30} \frac{d_A C_A^4}{(4\pi)^5} g^7 T^4   
 \;, \la{sc_est}
\ee
where we inserted $C_A = 3$ and $m_3 \sim gT$. In contrast, 
the effect that we have determined numerically in this paper
has the magnitude (cf.\ \eqs\nr{p_soft_splitup}, \nr{calFR})
\be
 \delta p_\rmi{soft} \sim 
 2 c_1 \frac{g_3^8 T}{m_3}
 \sim 20\, \frac{d_A C_A^4}{(4\pi)^5} g^7 T^4  
 \;.
\ee
This is larger than \eq\nr{sc_est} by more than 
two orders of magnitude. Another effect of $\rmO(g^7)$
is that the parameter $y$ should be computed at the 3-loop 
order; however, this only modifies the relation of $y$
and $T/\Tc$, and cannot revert the upward trend of our 
data at small $T/\Tc$. Therefore, it seems unlikely that 
infrared sensitive effects (from momenta $k\sim gT$)
describable with simple improvements 
of EQCD would be the cause for the mismatch; the problems
are perhaps more likely related to the treatment
of the ultraviolet modes ($k\sim \pi T$). 


%
\section{Conclusions and outlook}
\la{se:conclusions}

The purpose of this paper has been to approximate 
the non-perturbative contribution of the dynamics represented 
by a (truncated) effective theory called EQCD, \eq\nr{eqcd}, 
to the pressure of hot QCD. The result is constituted 
by the sum of the non-perturbative Linde term (\eq\nr{F_splitup}), 
perturbatively known terms up to 4-loop level
(\eqs\nr{p_soft_splitup}, \nr{calFxy}),
as well as an ultraviolet finite all-orders remainder that we have 
estimated numerically (\eq\nr{calFR}).

On the technical side, the main content of our study was to 
carefully carry out the continuum extrapolation needed for 
estimating the remainder, \eq\nr{calFR}. As has been illustrated 
in \fig\ref{fig:intercept}, the continuum extrapolation  
appears now to be under control, thanks partly to the recent
determination of the functions $f_i$ in \eq\nr{fit2}~\cite{nspt_a02}. 
Therefore, as we have argued, the representation in \eq\nr{calFR} should 
be free of substantial systematic errors close to the physical stripe
of \fig\ref{fig:stripe}. Comparing with the lower order terms 
in \eq\nr{calFxy}, \eq\nr{calFR} also has a reasonable magnitude, and
indicates that perturbation theory within EQCD is in fact a useful 
tool at all parameter values corresponding to the 4d theory. 
In principle, the numerical 
result could also be compared with improved resummation 
methods defined within EQCD (see, e.g., refs.~\cite{bir}--\cite{ck}). 

At the same time, from the phenomenological point of view, 
our result is somewhat of a disappointment: as shown in 
\fig\ref{fig:pheneos}, the match to 4d lattice data in the 
range $T/\Tc \ge 3.0$ is not particularly smooth. In fact, 
as shown in the figure, including the newly determined $\ge$ 5-loop 
remainder decreases the quality of the fit significantly. 
This implies that, unfortunately, we are not in a position to 
realize our original goal of offering consolidated 
crosschecks for the $\Nf\neq 0$ QCD pressure in the 
interesting temperature range $T/\Tc \sim 1.5 \ldots 3.0$. 

On the other hand, our study raises the theoretical question of
what kind of effects could be responsible for the mismatch between
our results and that of 4d lattice simulations. On the mundane
side, one possibility would be a substantial contribution from the 
condensate denoted by $\partial_x \mathcal{F}_\tinymsbar = 
 \langle (\tr[\hat A_0^2])^2 \rangle_a - \;...\;$. 
Unfortunately its systematic inclusion would require a significant
amount of new analytic and numerical work; moreover, as order-of-magnitude 
estimates and previous preliminary simulations suggest, it appears unlikely 
that this condensate could significantly change the qualitative 
behaviour that we have observed. 

On a more adventurous note, let us point out 
that, qualitatively, the reason for the mismatch is that the 
condensate in \eq\nr{A0R_fit} is {\em too large}, and grows rapidly 
as $y$ (or $T$) decreases
(note that for $\Nf=0$, the range $T/\Tc = 10^0 ... 10^3$
corresponds to $y \simeq 0.3 ... 1.2$, 
or $1/y \simeq 3.3 ... 0.8$, cf.\ \fig\ref{fig:stripe}). 
Since the condensate measures the mean
squared fluctuation of the field $A_0$, this means that the colour-electric
gauge field fluctuates too much. 
Indeed, if the remainder in \fig\ref{fig:intercept} continued to 
decrease also for $1/y\gsim 0.5$ (in principle the remainder can even 
become negative), then the match in \fig\ref{fig:pheneos} could 
conceivably improve dramatically. 
Perhaps one reason for the large fluctuations at small $y$ could be 
that the physical phase of EQCD is merely metastable~\cite{adjoint}
(see also Fig.~3 of ref.~\cite{su3adj})? 
If so, improved effective theories of the type 
in refs.~\cite{Z3}--\cite{cka}, 
which do not make any explicit distinction between 
the scales $k\sim gT$ and $k\sim \pi T$ and 
which are by construction free of the metastability problem, 
might yield smaller fluctuations and a better outcome. 

Of course, if no scale separation is made between the scales
$k\sim gT$ and $k\sim \pi T$, then one should in principle also 
account for corrections of $\rmO(g^8)$ to the hard part of 
the pressure ($p_\rmi{hard}$, \eq\nr{pQCD}), rendering the analysis
very hard. Nevertheless, it might be interesting to explore if 
a phenomenologically successful recipe could be found even without
this last step.  

%
\section*{Acknowledgements}

This work was partly supported by the Magnus Ehrnrooth Foundation, 
a Marie Curie Fellowship for Early Stage Researchers, and 
the Academy of Finland, contracts no.\ 114371, 122079.  
Simulations were carried out at the Finnish IT Center for Science (CSC). 
The total amount of computing time corresponds to about $1.4 \times 10^{18}$
flop.


\appendix
\renewcommand{\thesection}{Appendix~\Alph{section}}
\renewcommand{\thesubsection}{\Alph{section}.\arabic{subsection}}
\renewcommand{\theequation}{\Alph{section}.\arabic{equation}}


\section{Four-loop three-dimensional vacuum energy density}

The function 
$\mathcal{F}_\tinymsbar^\rmi{4-loop}$, 
defined through \eq\nr{F_splitup} and representing the contribution
of the adjoint scalar $A_0$ to the vacuum energy density of EQCD,
with $\bmu \equiv g_3^2$ and with $\Nc = 3$ so that only a single
scalar self-coupling appears, can be written as~\cite{aminusb}%
\footnote{%
 We have inserted the 
 value $\gamma_{10} = \pi^2/24-\ln^2\! 2/2$~\cite{ys}, 
 not known analytically in ref.~\cite{aminusb}.}: 
\ba
 \mathcal{F}_\tinymsbar^\rmi{4-loop}(x,y) & = & 
 - \frac{d_A}{(4\pi)} \frac{y^{\fr32}}{3}
 \nn & + & 
 \frac{d_A}{(4\pi)^2} \frac{y}{4} \biggl\{
 C_A \biggl[ 3 - 2 \ln(4 y)\biggr] + x (d_A+2) 
 \biggr\}
 \nn & + & 
 \frac{d_A}{(4\pi)^3} \frac{y^{\fr12}}{2} \biggl\{
 C_A^2 \biggl[ \frac{89}{12} + \frac{\pi^2}{3} - \frac{11}{3} \ln 2 \biggr]
 + \fr{x}2 C_A (d_A+2) \biggl[ 2 \ln(4 y) - 1\biggr]  
 \nn & & 
 + x^2 (d_A + 2)  \biggl[3 -  \ln(16 y)\biggr]
 - \fr{x^2}4 (d_A + 2)^2 
 \biggr\}
 \nn & + & 
 \frac{d_A}{(4\pi)^4} \biggl\{
 C_A^3 \biggl[
 \biggl(
 \frac{43}{8} - \frac{491}{1536} \pi^2  
 \biggr) \ln(4 y)
       + \frac{311}{256}  
       + \frac{43}{32} \ln 2
 \nn&&{} ~~~~ 
       + \frac{11}{3}\ln^2 \! 2 
       - \frac{461}{9216} \pi^2 
       + \frac{491}{1536} \pi^2 \ln 2 
       - \frac{1793}{512} \zeta(3)
 \biggr] 
 \nn & & 
 + \fr{5x}2 C_A^2 \biggl[ 
 \biggl( 1  - \frac{\pi^2}{8} \biggr) \ln(4 y)
        + 1 
        + 3 \ln 2 
        - \frac{17}{96}\pi^2 
        + \frac{\pi^2}{8} \ln 2
        - \frac{35}{16}\zeta(3)
 \biggr]
 \nn & & 
 + \fr{x}4 C_A^2 (d_A+2)  \biggl[ 
 \ln^2(4 y) - \ln(4 y) - \frac{43}{6} + \frac{11}{3} \ln 2 - \frac{\pi^2}{3} 
 \biggr]
 \nn & & 
 + \fr{x^2}4 C_A (d_A+2) \biggl[ 
 - \ln^2(4 y) + 
 \biggl( 
 9 - \frac{\pi^2}{4}
 \biggr) \ln(4 y)
 \nn&&{} ~~~~ 
 - 8 + 6 \ln 2 + \frac{31}{24}\pi^2  + \frac{\pi^2}{4} \ln 2 + 8 \ln^2 2
 - \frac{21}{4} \zeta(3)
 \biggr]
 \nn & & 
 - \fr{x^2}8 C_A (d_A+2)^2 \biggl[ 
 2 \ln(4 y)  + 1 
 \biggr]
 \nn & & 
 + \fr{x^3}4 (d_A + 2)^2   \biggl[ 
 \ln(16 y) - 1 
 \biggr] 
 + \frac{x^3}{24} (d_A + 2)^3  
 \nn & & 
 + \fr{x^3}{48} (d_A + 2)(d_A + 8)\biggl[ 
 - \pi^2 \ln(2 y) + \frac{\pi^2}{2} - 21 \zeta(3) 
 \biggr]
 \biggr\}
 \;. \la{calFxy}
\ea


\end{document}